\documentclass[12pt]{article}
\usepackage[dvips]{epsfig}
\usepackage{cite}
\newcommand {\Ordal}[1]{{O}(\alpha^{#1})}

\newcommand {\Placzek}{P\l aczek}
\newcommand {\Was}{W\c as}
\begin{document}
\title{PRECISION ELECTRO-WEAK AND HADRONIC LUMINOSITY CALCULATIONS}
\author{SCOTT A. YOST, CHRIS GLOSSER, and B.F.L. WARD\\
\it Department of Physics, Baylor University\\
\it P.O. Box 97316, Waco, TX 76798-7316\\[10pt]
\small E-mail:
Scott\_Yost@baylor.edu, Chris\_Glosser@baylor.edu, BFL\_Ward@baylor.edu}
\date{Presented at the\\ {\it Third International Symposium on Quantum Theory and Symmetries}\\ Cincinnati, September 2003\\[10pt] \sf BU-HEPP-03/13
}
\maketitle
\thispagestyle{empty}
\begin{abstract}
We have used YFS Monte Carlo techniques to obtain per-mil level
accuracy for the Bhabha scattering cross section used in the luminosity
monitor in electro-weak scattering experiments. We will describe
techniques for extending these methods for use in the $W$ production
luminosity cross section for hadron colliders.
\end{abstract}
\newpage 
\section{Introduction}

Following the discovery of the $W$ and $Z$ bosons, rapid
progress was made in precision measurements of electro-weak physics. 
By the end of LEP's run, high precision $Z$ data reached the $0.1\%$ level, 
creating pressure for the theoretical calculation of all relevant processes 
to reach the same level.

The beam luminosity enters into
all quantities requiring a normalized cross section, so its precise
measurement and calculation are crucial.  In $e^+ e^-$ scattering, the
luminosity is calibrated using small-angle Bhabha scattering, 
\hbox{$e^+ e^- \rightarrow e^+ e^- + n\gamma$}. This process has 
both experimental and theoretical advantages: it has a large, clean
signal and is almost pure QED, with only a $3\%$ contribution from 
$Z$ exchange at LEP energies. 

The matrix element for small-angle Bhabha scattering was computed by
adding the required radiative corrections to reach the desired 
precision level, and incorporating the resulting matrix element
into a Monte Carlo (MC) generator, BHLUMI.\cite{bhlumi} 
The MC algorithm was designed to implement Yennie-Frautschi-Suura (YFS)
exponentiation,\cite{YFS} which rigorously cancels infrared divergences to
all orders. 

We will review the precision of BHLUMI, and describe the additions
to BHLUMI that will be required to go beyond current technology in the event
of the construction of proposed $e^+ e^-$ linear colliders, which will have
larger-angle acceptance for the luminosity monitor. We also describe
a proposal for extending the methods developed for precision electro-weak
measurements for use in the luminosity monitor for the LHC or other 
advanced hadron colliders, where $W$ production is a leading candidate
for the luminosity process. 

\section{Two Photon Contributions to the Bhabha Luminosity Process}

\begin{figure}[t]
\setlength{\unitlength}{0.1mm}
\centerline{\begin{picture}(1200,700)
\put(250,0){\makebox(0,0)[cb]{\sf\small (a)}}
\put(-200,80){\makebox(0,0)[lb]{
\epsfxsize=75mm\epsfbox{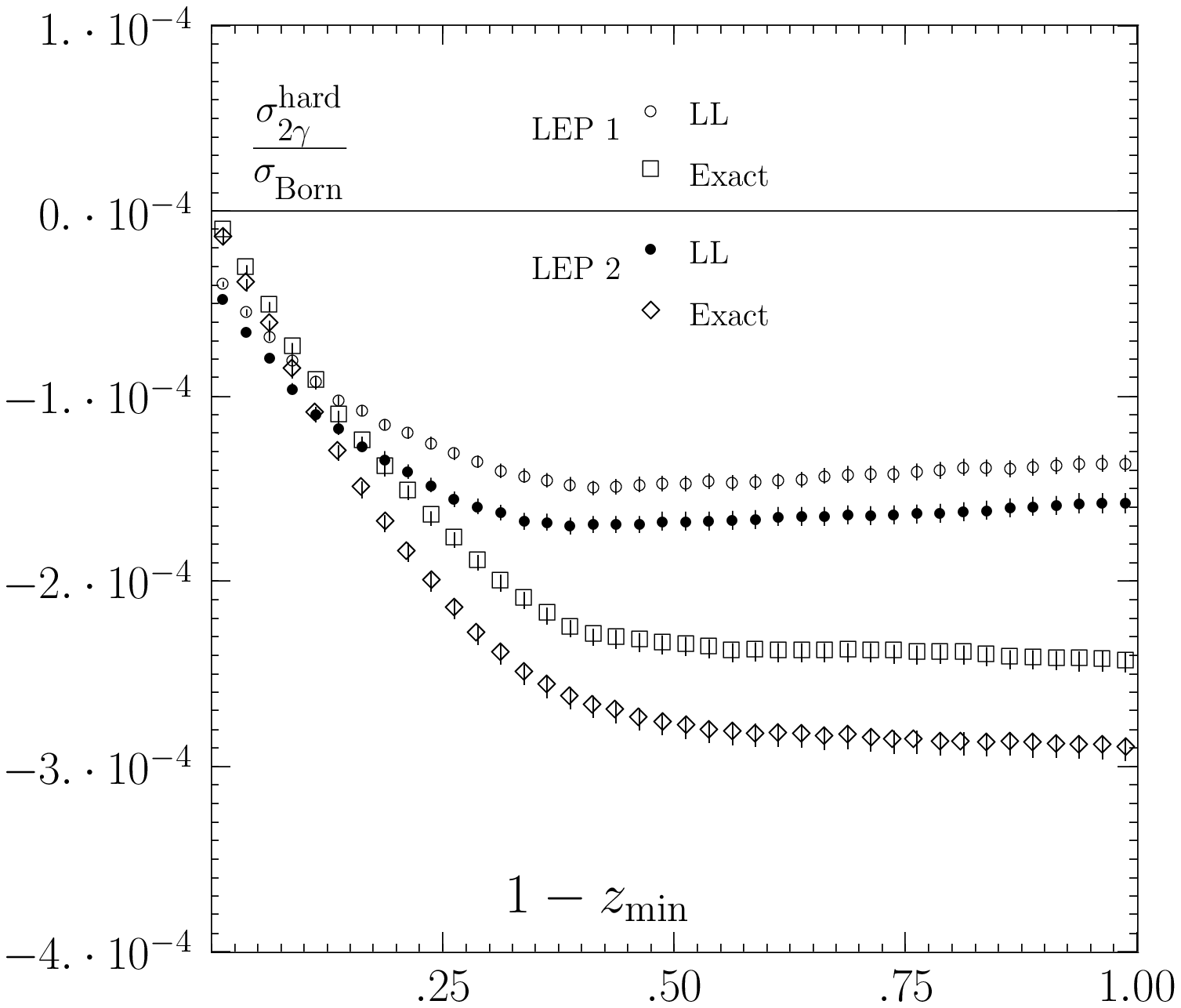}}
}
\put(600,180){\makebox(0,0)[lb]{
\epsfxsize=75mm\epsfbox{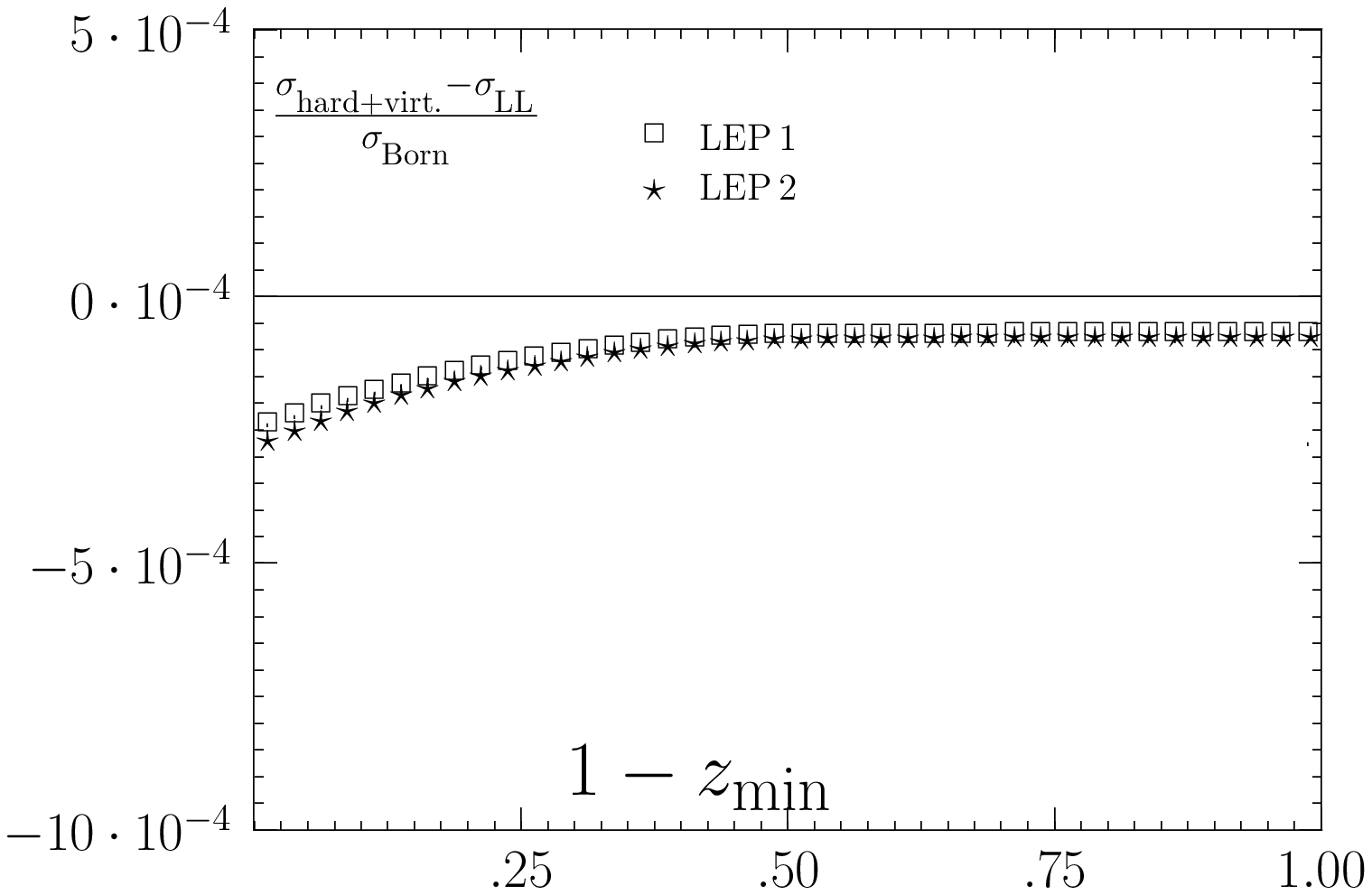}}
}
\put(1050,0){\makebox(0,0)[cb]{\sf\small (b)}}
\end{picture}}
\sf\small\caption{\sf\small Monte Carlo results ($10^6$ events) for the $\Ordal{2}$ cross 
section for (a) two hard photon emission and (b) single hard plus virtual 
photon emission from the electron line for LEP1 and LEP2 parameters, 
where $z_{\min}$ is a lower bound on the fraction of the beam energy 
carried away by the electron and positron.  The cross sections are 
normalized by dividing by the Born cross section, and in (b), the leading 
log contribution is subtracted. 
}
\label{fig:mc}
\end{figure}

 It was recognized that to reduce the error estimate of BHLUMI to the
per-mil level or better, it would be necessary to compute the exact 
two-photon radiative corrections, which previously had been incorporated in
a ``leading log'' (LL) approximation.  The first step was to
calculate exactly the cross section for emitting two hard photons.\cite{2hard} 
The LL and exact results are compared in
Fig. \ref{fig:mc}(a) for LEP1 parameters (beam energy 91 GeV, angles 
between $1^\circ$ and $3^\circ$) and LEP2 parameters (beam energy 176 GeV, 
angles between $3^\circ$ and $6^\circ$). It is seen that the leading log 
result was accurate to within $0.013\%$ in both cases.

\begin{figure}[ht]
\centerline{\epsfxsize=4.0in\epsfbox{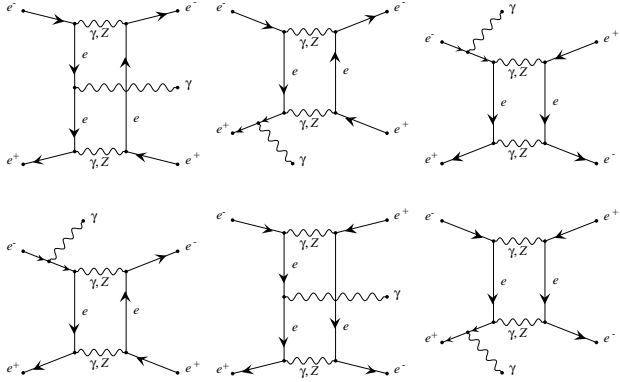}}
\sf\small\caption{\sf\small Box diagrams contributing to Bhabha scattering radiative corrections 
at larger angles. The $s$ channel diagrams are shown. The crossed
versions are required for the $t$ channel.}
\label{fig:boxes}
\end{figure}
Subsequently, mixed hard and virtual photon correction to Bhabha scattering
were calculated exactly in the small-angle regime.\cite{hard+virtual} 
All relevant diagrams were included except for the ``box diagrams''  shown in 
Fig.\ \ref{fig:boxes}, which become significant only at larger angles.
The difference between the exact result and the leading log result 
implemented in BHLUMI is shown in Fig.\ \ref{fig:mc}(b) for both LEP1 
and LEP2 parameters.
In the experimentally interesting range $0.2 \le 1 - z_{\rm min} \le 1.0$,
BHLUMI is within $0.02\%$ of the exact result for both LEP1 and LEP2. 

The second-order photonic corrections were completed by adding the two-loop
virtual photon correction to Bhabha scattering from Ref.\ \citen{2loop}, which
yielded a $0.014\%$ contribution to the cross section.\cite{vancouver}
The combined contribution of the missing order $\alpha^2$ photonic radiative
corrections in BHLUMI turned out to be $0.027\%$. The final BHLUMI precision
tag was reduced to $0.061\%$ for LEP1 parameters, and to $0.122\%$ for 
LEP2 parameters.

\section{Bhabha Luminosity for Linear Colliders}
The luminosity monitors for proposed linear $e^+ e^-$ colliders will have larger
angle acceptances, which requires extending the exact low-angle 
second order photonic corrections beyond the small-angle regime. 
The box diagrams in Fig.\ \ref{fig:boxes} that were neglected 
in the previous calculation must be added in this case. 

Calculating the box diagrams requires one new ingredient not needed for
the previous calculations: a five-point off-shell box diagram. An algorithm
for this diagram is currently under construction. When complete, the addition
of these box diagrams will complete the exact order $\alpha^2$
photonic contribution to Bhabha scattering. 

\section{Hadronic Luminosity Monitor}

The proposed $W$-production luminosity process at the LHC will 
require at least a $1\%$ precision
level for the theoretical contribution to the data analysis. Reducing the
theoretical uncertainty to this level will require all first and second
order QCD radiative corrections, as well as first-order electro-weak radiative
corrections, and mixed QCD -- electro-weak corrections. Due to the large
number of graphs, automated techniques are essential. Those displayed 
in this paper were are excerpted from the output of GRACE.\cite{grace}
                                                                                
The first-order electro-weak radiative corrections to $u{\overline d}
\rightarrow W$ consist of three real photon emission graphs and 19 graphs
including a virtual photon or $Z$.  These can be computed with well-known
methods. The first order gluonic corrections are likewise known, or can
be calculated using well-known techniques. 

\begin{figure}[ht]
\centerline{\epsfxsize=4.0in\epsfbox{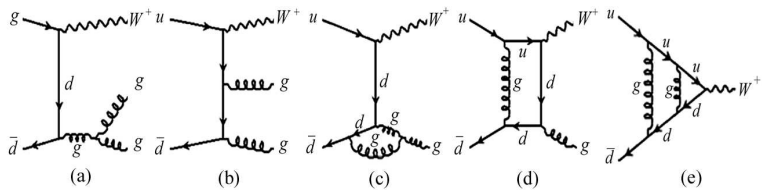}}
\sf\small\caption{\sf\small Representative two-gluon diagrams: (a,b) two real gluons, 
(c,d) real + virtual gluon, (e) two virtual gluons.} 
\label{fig:2gluon}
\end{figure}

The two-gluon radiative corrections are the closest analog to the calculations
that were needed in the $e^+e^-$ luminosity case. These are complicated 
by the triple-gluon coupling in the QCD case, however. 
Fig.\ \ref{fig:2gluon} shows some representative examples of 
the relevant graphs.  There are eight graphs for emitting two
real gluons, 13 mixed real + virtual gluon (one loop) graphs,
and 22 graphs with two virtual gluons (two loop) graphs.
The latter clearly present the greatest technical challenges. 
All of these results will be needed
to NLL order to reach the $1\%$ precision level. Thus, they will contribute
an ${O}(\alpha^2_s L)$ term to the cross section, with $L$ a typical
``big logarithm'' for the calculation.

\begin{figure}[ht]
\centerline{\epsfxsize=4.0in\epsfbox{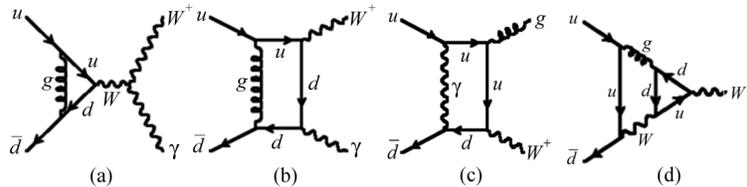}}
\sf\small\caption{\sf\small Representative mixed hadronic -- electro-weak diagrams:
(a,b) real photon + virtual gluon, (c) real gluon + virtual photon,
(d) virtual gluon + electro-weak loop.} 
\label{fig:mixed}
\end{figure}

The next corrections will
be mixed strong and electro-weak radiative corrections, including the
representative graphs shown in Fig.\ \ref{fig:mixed}. There are 10
graphs with a virtual gluon and real photon emission, 86 graphs with a 
virtual photon or $Z$ and real gluon emission, and 293
two-loop graphs with a virtual gluon and electro-weak loop. 

Pure second-order electro-weak radiative corrections will be needed as well,
but only to leading log order, adding a ${O}(\alpha^2 L^2)$ contribution.
The matrix elements will be combined with DGLAP evolved structure
functions\cite{dglap} and incorporated into a MC program.  Progress on a 
precision calculation ($0.1\%$) of the structure function evolution 
has recently been reported using MC methods.\cite{markov} 

An important aspect of
BHLUMI's success was the YFS exponentiation, which permitted an
exact cancelation of all infrared singularities to all orders. We expect
YFS exponentiation to play an important role in the hadronic MC as well. Some
relevant techniques have already been developed for QCD processes,
originally motivated by anticipation of the SSC.\cite{sscyfs}

\section{Conclusions}

We have reviewed the progress which led the electro-weak 
Bhabha luminosity process to the per-mil precision level and beyond. 
Verifying this precision required exact calcluations of all 
second-order photonic radiative corrections to small angle Bhabha scattering.
A few ``box diagrams,'' which become important at larger-angle scattering,
are still in the process of being calculated. Adding
these box diagrams will bring to completion a 12-year project to compute
all of these processes.

The construction of the LHC and other next-generation colliders will soon place
unprecedented precision requirements on the calculations of the hadronic
and electro-weak processes measured at those colliders. A luminosity
process calculation on the order of $1\%$ will
be needed to fully test the validity of the Standard Model, and to search 
effectively for hints of new physics.

\section*{Acknowledgments}

S.Y. thanks the conference organizers for an invitation to present this
paper, and Baylor University for providing funding to attend the conference.
This work was supported in part by by the US Department of Energy contract  
DE-FG05-91ER40627 and by NATO Grant PST.CLG.977751.

\end{document}